\documentclass[10pt,a4paper]{spie}
\usepackage[utf8]{inputenc}
\usepackage[dvipdf]{graphicx}
\usepackage{makeidx}
\usepackage{color}
\usepackage{multicol}
\usepackage{subfig}

\title{
   Tilt angle measurement with a Gaussian-shaped laser beam tracking   
    }

\author{
	Martin \v{S}arbort,
    \v{S}imon \v{R}e\v{r}ucha,
    Petr Jedli\v{c}ka,
	Josef Lazar,
	Ond\v{r}ej \v{C}\'{i}p
\skiplinehalf
Institute of Scientific Instruments, AS CR (ISI) , Kr\'{a}lovopolsk\'{a} 147, 612 64 Brno, Czech Republic
\skiplinehalf
}

\authorinfo{Further author information: please send correspondence to e-mail address martins@isibrno.cz (M. Sarbort)}

\begin{document}
\maketitle


\begin{abstract}
We have addressed the challenge to carry out the angular tilt stabilization of a laser guiding mirror which is intended to route a laser beam with a high energy density. Such an application requires good angular accuracy as well as large operating range, long term stability and absolute positioning. We have designed an instrument for such a high precision angular tilt measurement based on a triangulation method where a laser beam with Gaussian profile is reflected off the stabilized mirror and detected by an image sensor. As the angular deflection of the mirror causes a change of the beam spot position, the principal task is to measure the position on the image chip surface. We have employed a numerical analysis of the Gaussian intensity pattern which uses the nonlinear regression algorithm. The feasibility and performance of the method were tested by numeric modeling as well as experimentally. The experimental results indicate that the assembled instrument achieves a measurement error of 0.13 microradian in the range $\pm$0.65 degrees over the period of one hour. This corresponds to the dynamic range of 1:170\,000.
\end{abstract}

\keywords{optical alignment, tilt angle measurement, laser beam tracking, metrology}\\

{\bf DOI:} 10.1117/12.2052880 \\

Note: This is a preprint rendition of a conference paper M. Sarbort et al,  Tilt angle measurement with a Gaussian-shaped laser beam tracking, Proc. SPIE, vol. 9132, pp. 91321E, 2014.
%
%
%
%

\section{Introduction}

One of the most challenging scientific projects currently ongoing in Europe is Extreme Light Infrastructure (ELI) which is part of a European plan to build a new generation of large research facilities. The first research center ELI Beamlines is located in the Czech Republic. It aims to develop an internationally unique laser device that will provide platform for research and application projects involving the interaction of light with matter at intensity which is about ten times higher than the currently achievable values. ELI will provide ultra-short laser pulses of a few femtoseconds (10-15 fs) duration and give performance up to 10 PW. These extraordinary characteristics predispose ELI to bring new techniques for medical image-display and diagnostics, radiotherapy, tools for new materials developing and testing, latest in X-ray optics, etc.

The initial research program involves development and construction of the laser device itself, which represents a major challenge due to extreme scientific and technical demands. To attain high focused intensities intended for ELI it is necessary to utilize optical compressors and amplifiers that consist of many optical elements enclosed in vacuum chambers and interconnected by vacuum pipes through which the laser beam is routed. Given the size of these devices the path length of the laser beam reaches hundreds of meters. Therefore, an extremely precise positioning of involved optical elements becomes an important issue.

In this paper we present a method to measure the angular tilt of the laser guiding mirrors designed for ELI laser with desired angular accuracy of 1~microradian, operating range $\pm 4^{\circ}$, long term stability and absolute positioning in two axes. Our approach is based on a triangulation method where a laser beam profile is reflected off the stabilized mirror and detected by an image sensor. Since the angular deflection of the mirror causes a change of the beam spot position, the principal task is to measure the position on the image chip surface. For this purpose we have employed a numerical analysis of the Gaussian intensity pattern which uses the nonlinear regression algorithm. The performance of the method was tested numerically and verified experimentally with satisfactory results that exceed the required accuracy.

The rest of the paper is organized as follows: Section \ref{met} specifies the requirements imposed on the measurement system, describes the proposed method and estimates its performance by means of numerical testing. Section~\ref{exp} presents experimental verification of our method. Section \ref{res} holds the discussion and Section \ref{fin} concludes the paper.

\begin{figure}[th]
\centering
\includegraphics[width=.51\textwidth]{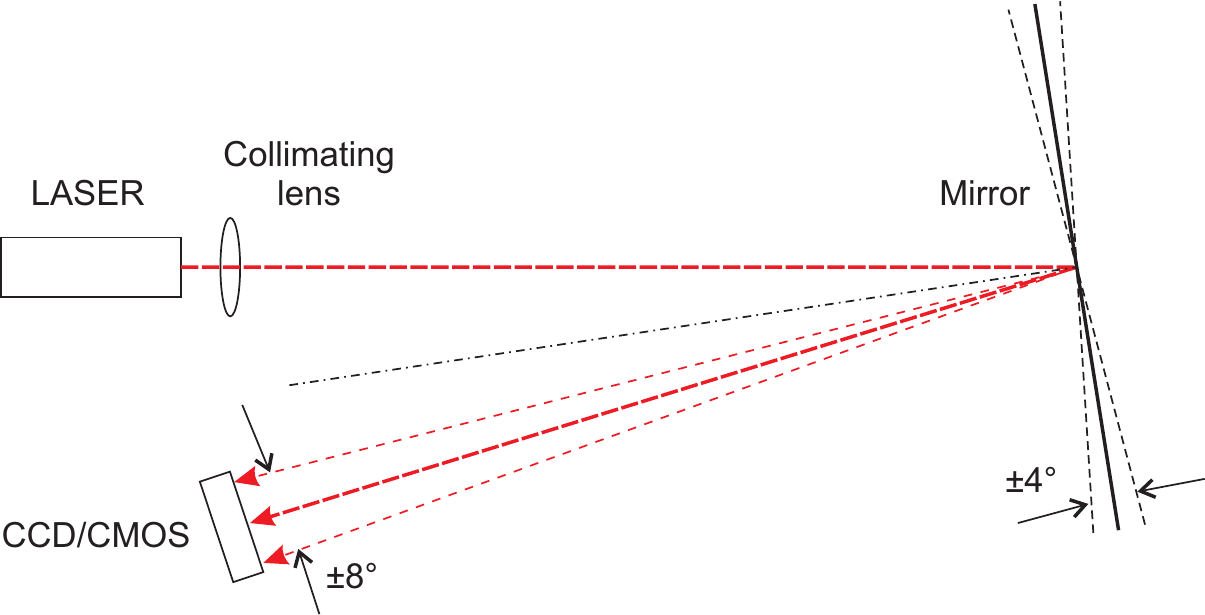}\vspace{1.5mm}
\caption{Principle of triangulation method.}
\label{f_triangulation}
\end{figure}


\section{Methods}
\label{met}

The requirements imposed on the tilt angle measurement system are given by high angular resolution of 1~microradian, large operating range $\pm 4^{\circ}$, long term stability in hours and absolute positioning in two perpendicular axes. The key problem was neither the high accuracy nor the large measurement range alone, but the combination of both these requirements together with demand for two-dimensional measurements. The target dynamic range is 1:150\,000. 

The preliminary considerations selected three possible approaches to the tilt angle measurement -- a triangulation method with laser beam tracking \cite{Hofherr2013,Rakich2012}, an autocollimator-based method \cite{Arp2013,Yuan2005} and an interferometric method based on differential measurements in two parallel laser beams \cite{Hrabina2011,Hahn2010}. The main disadvantage of the autocollimator is a limited dynamic range, the main problem of interferometric methods is the impossibility of absolute measurement. Considering these drawbacks we selected the triangulation as the preferred measurement approach. 

The principle of triangulation method is shown in Figure \ref{f_triangulation}. It utilizes a laser beam emerging from the source that is sequentially reflected off a mirror and detected by an image sensor. The angular deflection of the mirror is determined from the measured deflection of the laser spot position on an image sensor and the given geometrical configuration of the measurement system. Considering the measurement distance of 1~meter, the desired angular accuracy of 1~microradian corresponds to the laser spot detection with resolution of 1~micrometer. To achieve such a high accuracy it was necessary to choose proper hardware components of the measurement system (especially an image sensor) and to develop extremely precise numerical  method for detection of the laser spot position.

Among the available laser sources we preferred the helium-neon lasers that are relatively cheap and robust devices providing an output beam with a Gaussian intensity profile of excellent quality. As the most suitable image sensors we selected the full-frame CCD/CMOS chips that are characterized by high spatial resolution (pixel size about 5\,$\mathrm{\mu m}$), relatively large size ($28 \times 35$\,mm), standard bit depth (8-12 bits) and high sample rates (hundreds of frames per second). 

\begin{figure}[th]
\centering
\hspace{0.25cm}
\subfloat[]{\includegraphics[height=4.2cm]{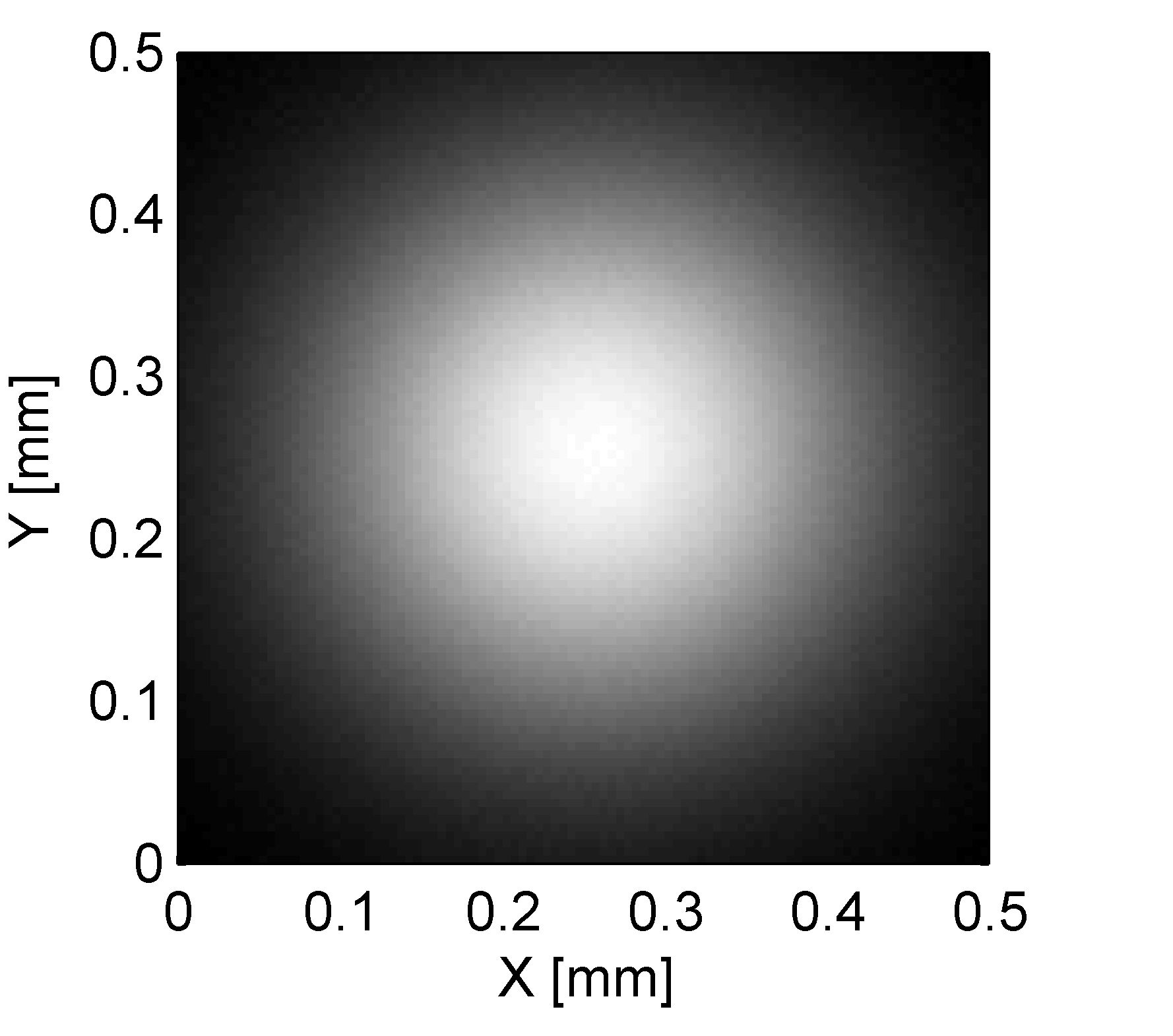}}\hspace{0.75cm}
\subfloat[]{\includegraphics[height=4.2cm]{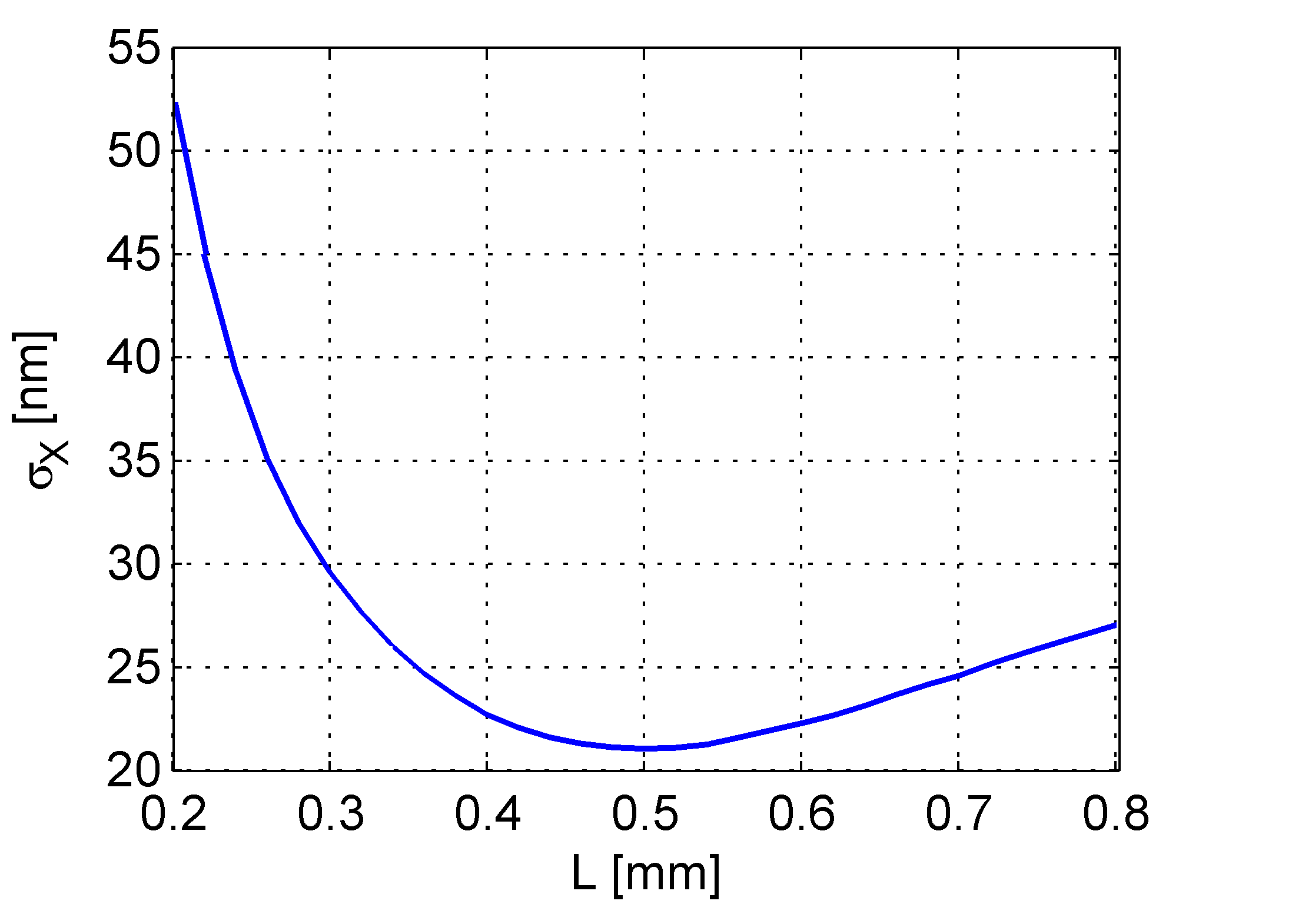}}\\
\vspace{-3mm}
\subfloat[]{\includegraphics[height=4.2cm]{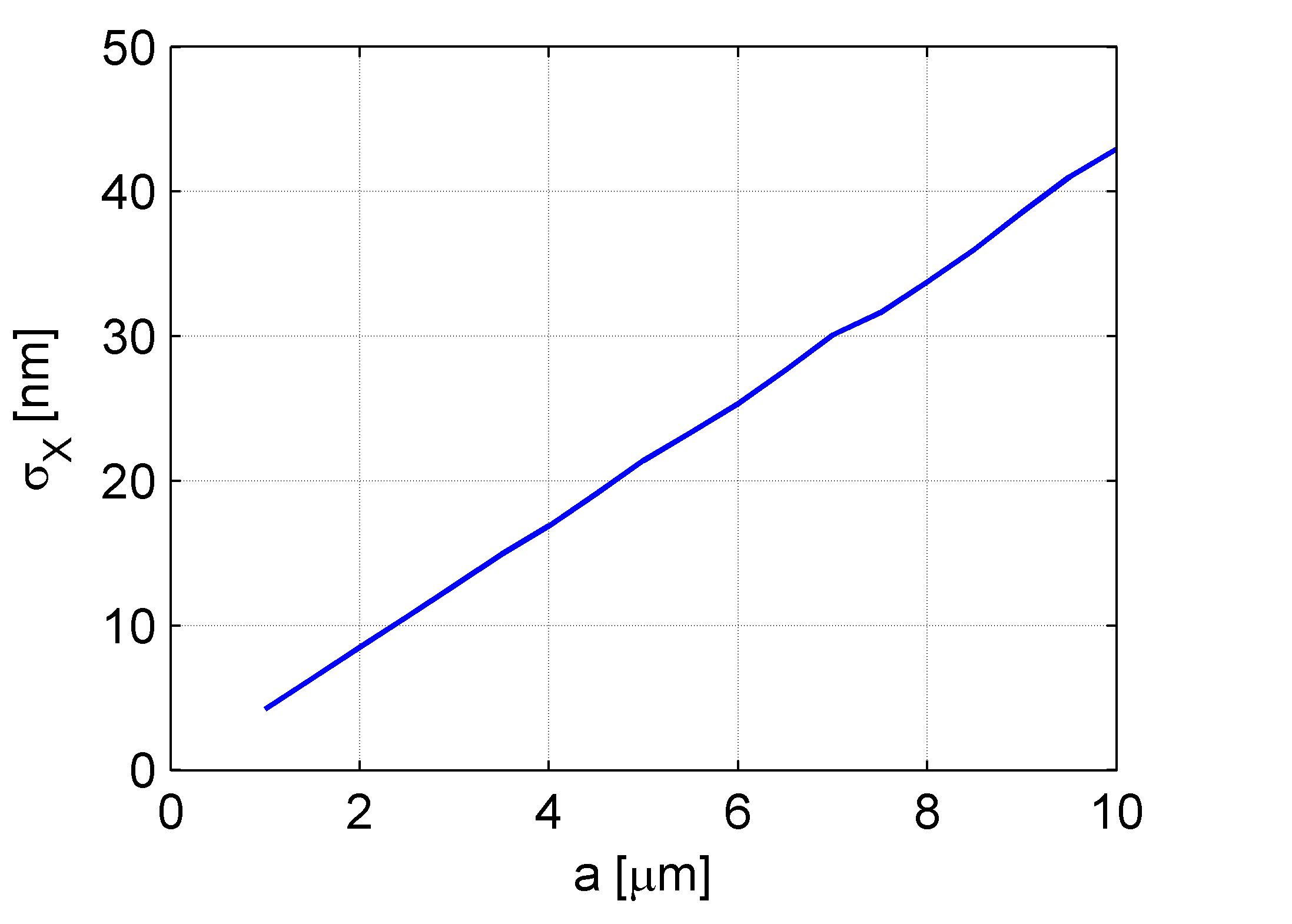}}
\subfloat[]{\includegraphics[height=4.2cm]{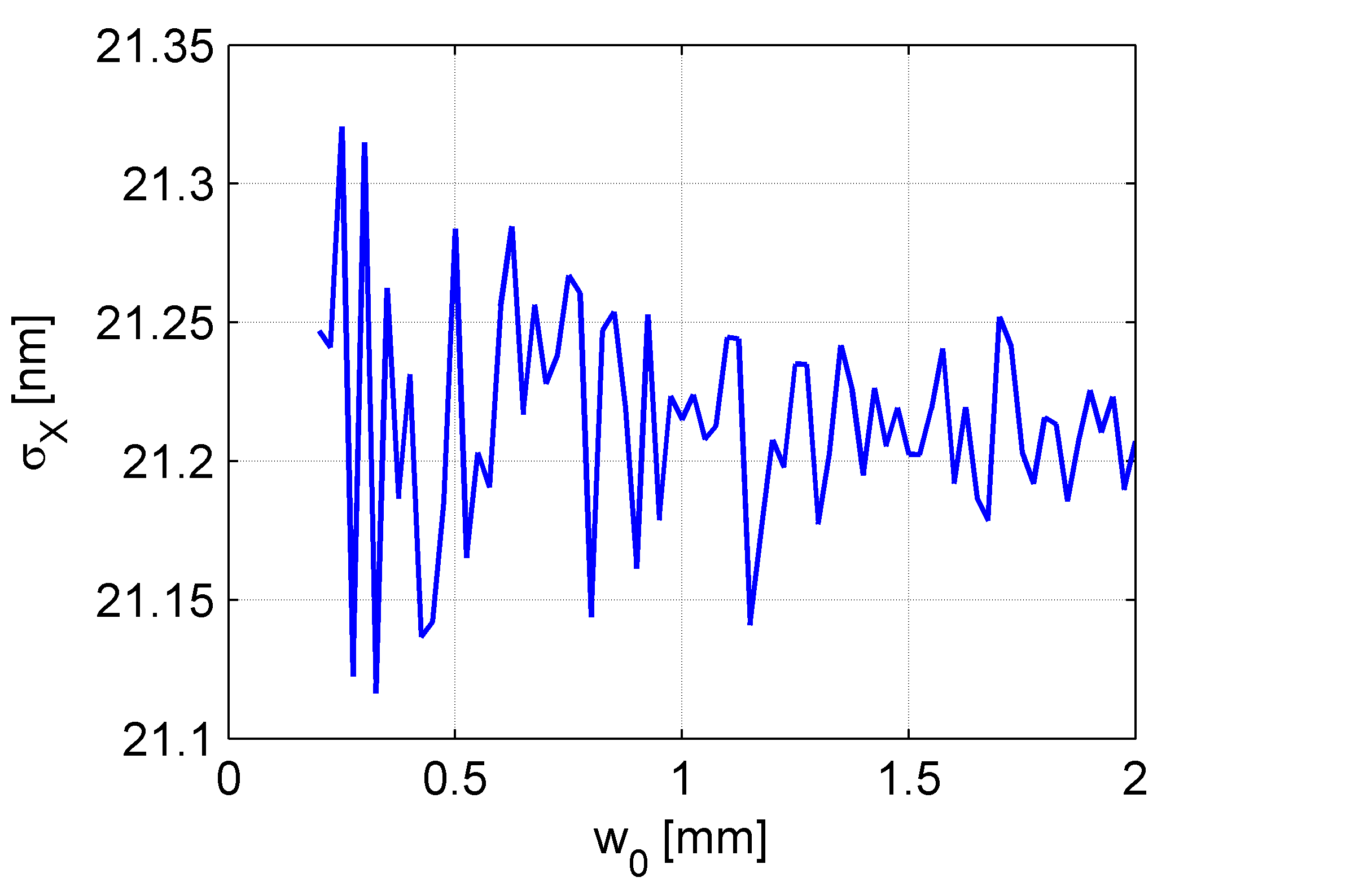}}
\caption{The results of the numerical simulation for a Gaussian laser beam. a) The intensity profile. The dependence of the error $\sigma_x$ on b) the side length $L$ of the square area used for nonlinear regression, c) the pixel size $a$ and d) the beam radius $w_0$.}
\label{f_simulation}
\end{figure}

The laser beam tracking on an image sensor lies in a numerical analysis of the detected intensity pattern  by means of non-linear regression. The key idea is that an expected intensity pattern spanning multiple pixels allows for detection of its position with resolution better than the size of a single pixel. The simplest case is to detect a single laser spot with the Gaussian intensity profile given in the Cartesian coordinates by 
\begin{equation}
I(X,Y) = I_0 \exp \left[ - \frac{2 \left( (X-X_0)^2 + (Y-Y_0)^2 \right) }{w_0^2} \right ] ,
\label{gauss_intensity}
\end{equation}
where $[X_0,Y_0]$ is the desired center point and $w_0$ is the beam radius. Due to the computational limitations only a part of the entire chip can be used for non-linear regression. Therefore the evaluation algorithm must be divided into two steps. First, a coarse position and size of the expected intensity pattern is detected and an appropriate square area of the side length $L$ is selected. Second, the Levenberg-Marquardt algorithm is used to fit an expected analytical model (\ref{gauss_intensity}) to the observed intensity pattern.  As an output we obtain the model parameters including the desired coordinates $X_0$, $Y_0$ and the corresponding errors $\sigma_X$, $\sigma_Y$ given at the standard confidence level 68.3$\%$.

The capabilities of the proposed method for laser beam tracking on the sensor surface were tested by means of numerical modeling. Assuming an image sensor with 8 bit resolution and the noise level of two least significant bits, we investigated the limits of the accuracy given by the errors $\sigma_X$ and $\sigma_Y$, the measurement range and the corresponding dynamic range. 

The accuracy assessment requires to consider the errors $\sigma_X$ and $\sigma_Y$ as the functions of the pixel size $a$, the beam radius $w_0$ and the side length $L$ of the square area used for nonlinear regression. In Figure \ref{f_simulation} we show the results of the numerical simulation for three different parameters settings where only one parameter is variable and the others are fixed:
\begin{itemize}
\item $w_0 = 0.25\,\mathrm{mm}$, $a = 5.0\,\mathrm{\mu m}$, $L \in [0.2, 0.8]\,\mathrm{mm}$,
\item $w_0 = 0.25\,\mathrm{mm}$, $a \in [1,10]\,\mathrm{\mu m}$, $L = 2 w_0$,
\item $w_0 \in [0.2, 2.0]\,\mathrm{mm}$, $a = 5.0\,\mathrm{\mu m}$, $L = 2 w_0$.
\end{itemize}
Apparently, the error $\sigma_X$ (and analogously $\sigma_Y$) depends on the side length $L$ nonlinearly. The minimal error $\sigma_X = 21.2$\,nm is obtained when $L$ is approximately equal to the beam diameter, i.e. $L \approx 2w_0$. This can be justified as follows -- if $L < 2w_0$ only a part of the beam profile is used for the nonlinear regression and, therefore, the error $\sigma_X$ becomes higher; if $L > 2w_0$, there is a large peripheral area where a signal-to-noise ratio is very small which results in increase of $\sigma_X$. Therefore, we chose the length $L = 2w_0$ as an optimal value for further investigations. Considering the fixed $w_0$, we found that $\sigma_X$ is approximately linearly proportional to $a$. Finally, we chose a fixed $a$ and considered the possible values of $w_0$ in the range from a few tenths of a millimeter up to several millimeters. It turned out that $\sigma_X$ varies only in the range of few tenths of a nanometer and, therefore, $\sigma_X$ can be considered as independent of $w_0$. 

\begin{figure}[t]
\centering
\subfloat[]{\includegraphics[height=4.2cm]{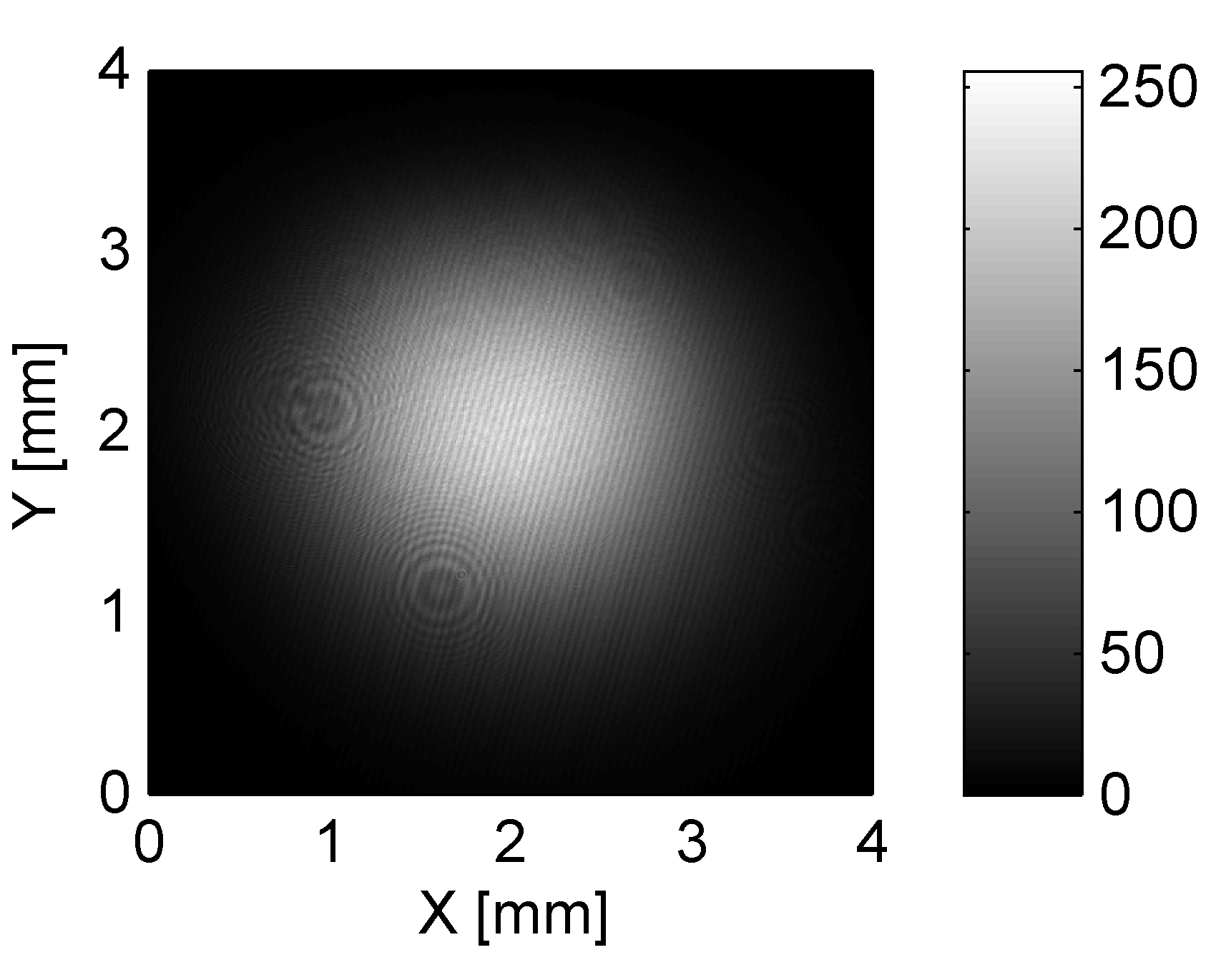}}
\hspace{3.5mm}
\subfloat[]{\includegraphics[height=4.2cm]{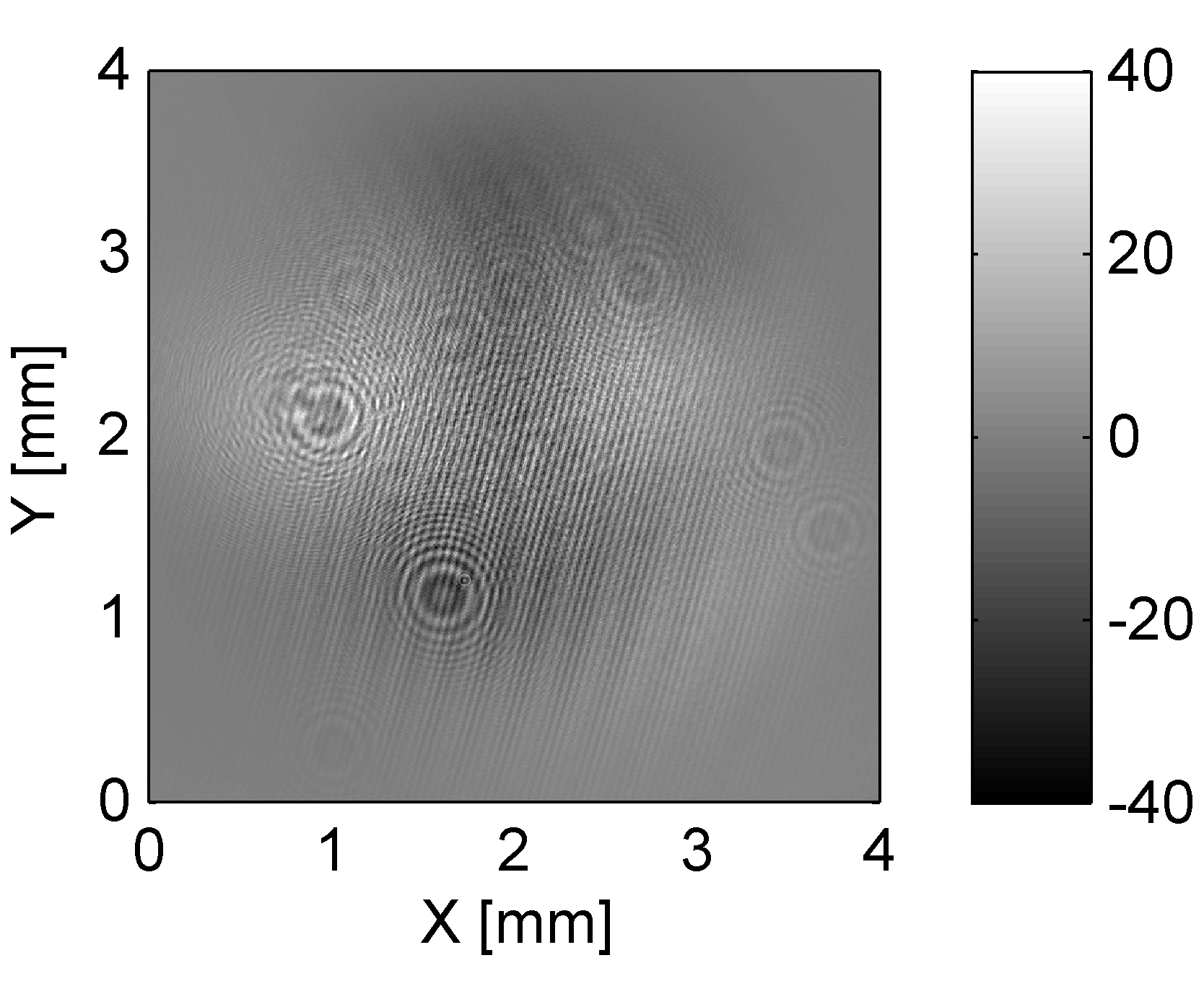}}
\hspace{3.5mm}
\subfloat[]{\includegraphics[height=4.2cm]{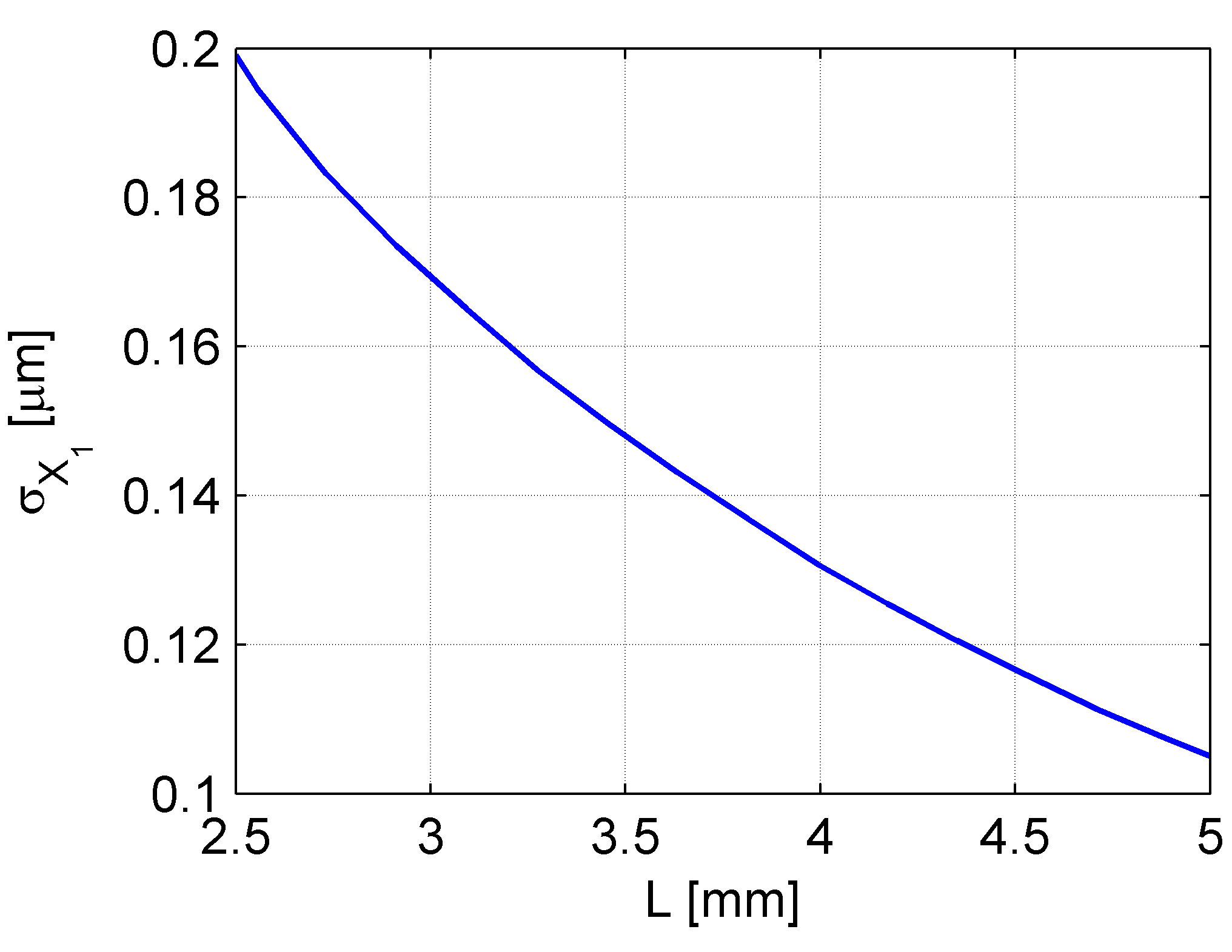}}
\caption{The results of a single measurement: a) Observed intensity $I_{\mathrm{obs}}$, b) difference between $I_{\mathrm{obs}}$ and the Gaussian profile $I_{\mathrm{reg}}$ calculated by nonlinear regression, c) the dependence of the error $\sigma_{X_1}$ on the length $L$. }
\label{f_experiment_single}
\end{figure}

The linear measurement range along a single axis of the image sensor is generally given by the difference of the image sensor size $A$ and the side length $L$ of the square area used for nonlinear regression. Since the full-frame sensor is rectangular in shape we take the smaller side length $A = 28$\,mm as the relevant value. 

Finally, we calculate the corresponding dynamic range given by the ratio $\sigma_X/(A-L)$. Considering the full-frame image sensor and taking the minimal error obtained from the above numerical simulation, we get the dynamic range approximately 1:1\,300\,000. This is almost one order of magnitude better than the desired value 1:150\,000. Therefore, based on the results of numerical simulations, we conclude that the triangulation method with laser beam tracking is suitable to achieve the demands imposed on the tilt angle measurement system.


\section{Experimental Results}
\label{exp}

The experimental setup assembled in order to verify capabilities of the triangulation method with laser beam tracking consisted of the laser source, the digital camera and the PC with the control and acquisition software. This simple setup allowed for identification and elimination of error sources that may considerably affect the measurement precision.

We have used a 0.8\,mW helium-neon laser source operating at 632.8\,nm. The output beam was characterized by  the waist radius 0.24\,mm and the divergence 1.7\,mrad. The measurement distance $d$  between the laser source and the image sensor was 1050\,mm where the beam radius was about 1.45\,mm. The Gaussian profile of the laser beam was detected by a monochrome CCD sensor with dimensions $12.8 \times 9.6$\,mm, $2330 \times 1750$ pixels, the pixel size 5.5\,$\mathrm{\mu m}$ and the resolution 8\,bit. The experimental setup was assembled on a vibration-damped optical table in an underground laboratory with air condition providing stabilization of temperature. In addition, the beam path was surrounded by a protective tube to reduce possible fluctuations of the refractive index of air. 

The measurement was running in a loop that consisted of three steps: the image acquisition using the camera, the image processing by means of nonlinear regression and the data logging for further analysis of the results. The period of a single measurement was about 3 seconds. To prove the long-term stability, we have conducted overnight measurements lasting about 12 hours. 

The experiment evaluation was aimed to calculate the total measurement uncertainty of the laser spot position, taking into account the individual and the long-term measurements. A single measurement is characterized by the errors $\sigma_{X_1}$ and $\sigma_{Y_1}$ obtained from the nonlinear regression algorithm, i.e. these are identical to the quantities $\sigma_{X}$ and $\sigma_{X}$ defined in Section \ref{met}. The long-term measurements were divided into hour intervals that served for further evaluation. Taking approximately 1200 measurements recorded within one hour, we calculated the mean coordinates $\bar{X}_0$, $\bar{Y}_0$ and the corresponding standard deviations $\sigma_{X_2}$ and $\sigma_{Y_2}$. Then the total measurement uncertainty of the laser spot position is given by the combined errors $\sigma_{X_C}$ and $\sigma_{Y_C}$ defined by $\sigma_{X_C} = \sqrt{\sigma_{X_1}^2 + \sigma_{X_2}^2}$ (and analogously $\sigma_{Y_C}$).

\begin{figure}[t]
\centering
\includegraphics[width=.65\textwidth]{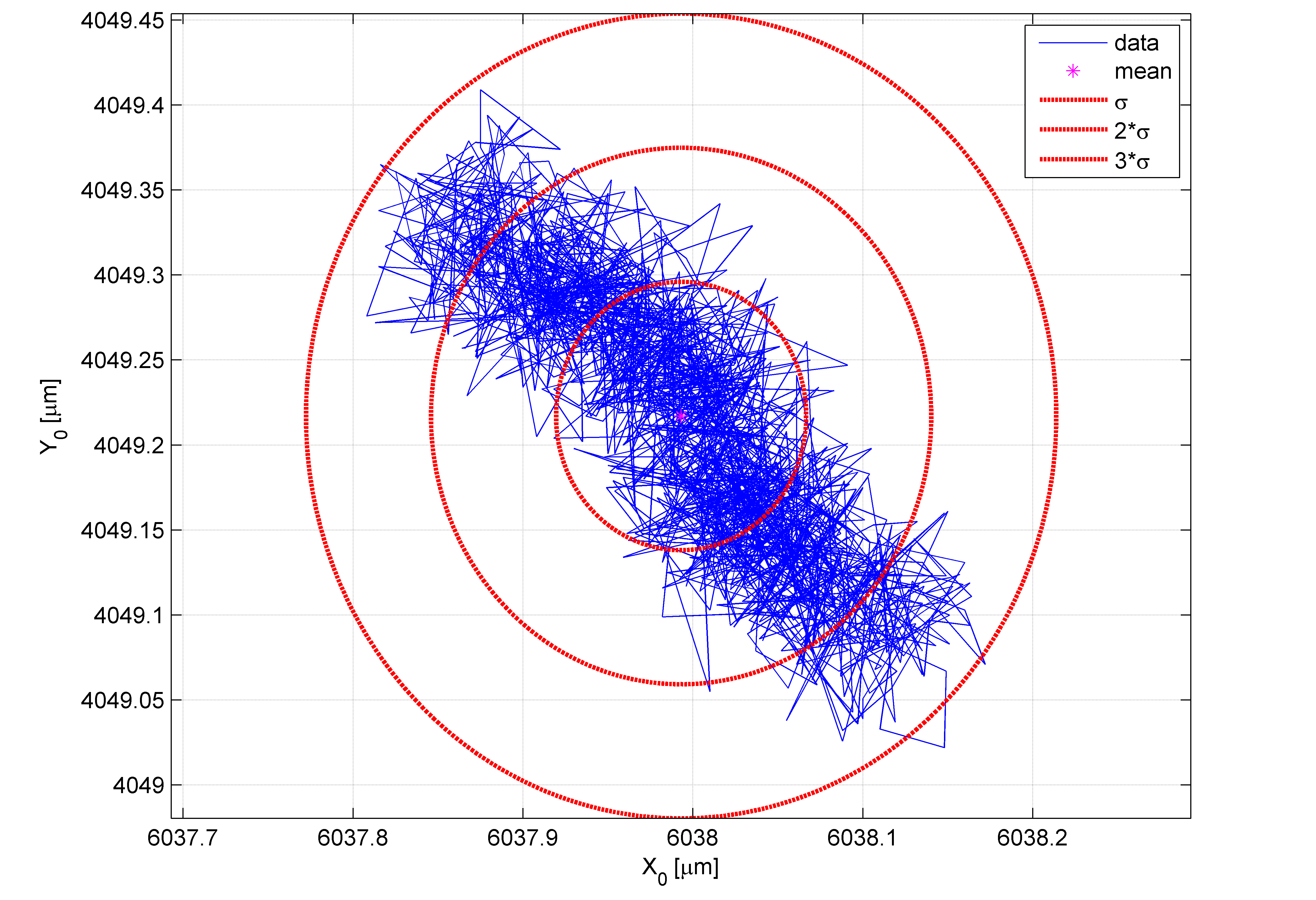}\\
\includegraphics[width=.65\textwidth]{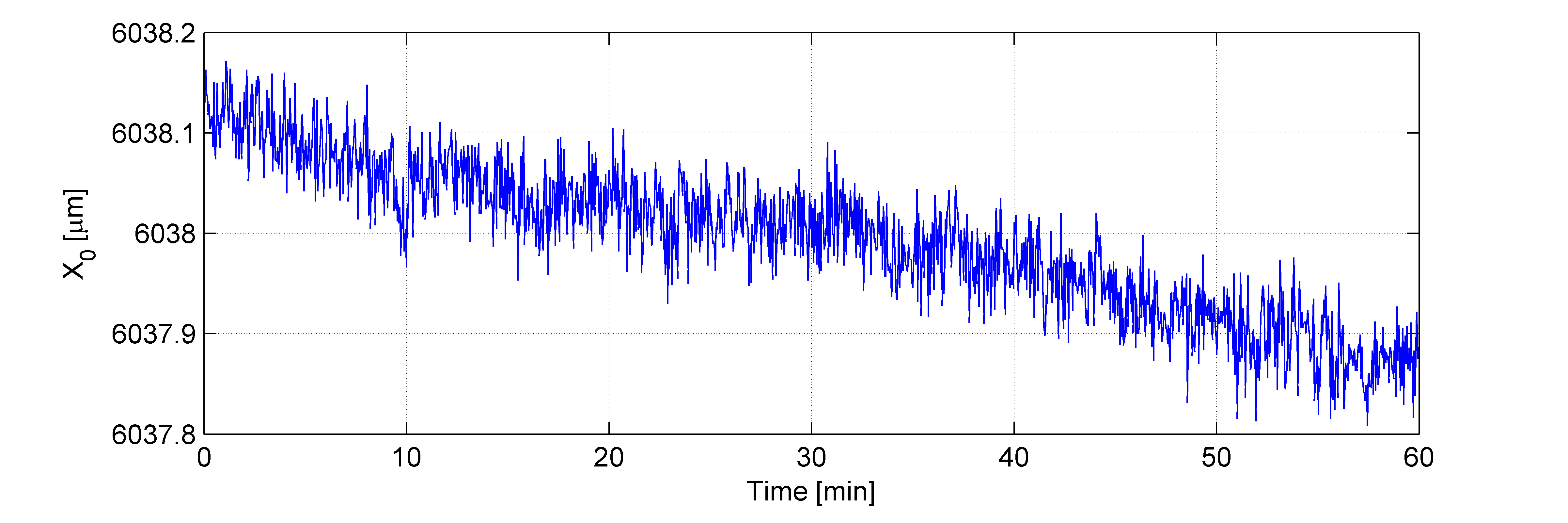}\\
\includegraphics[width=.65\textwidth]{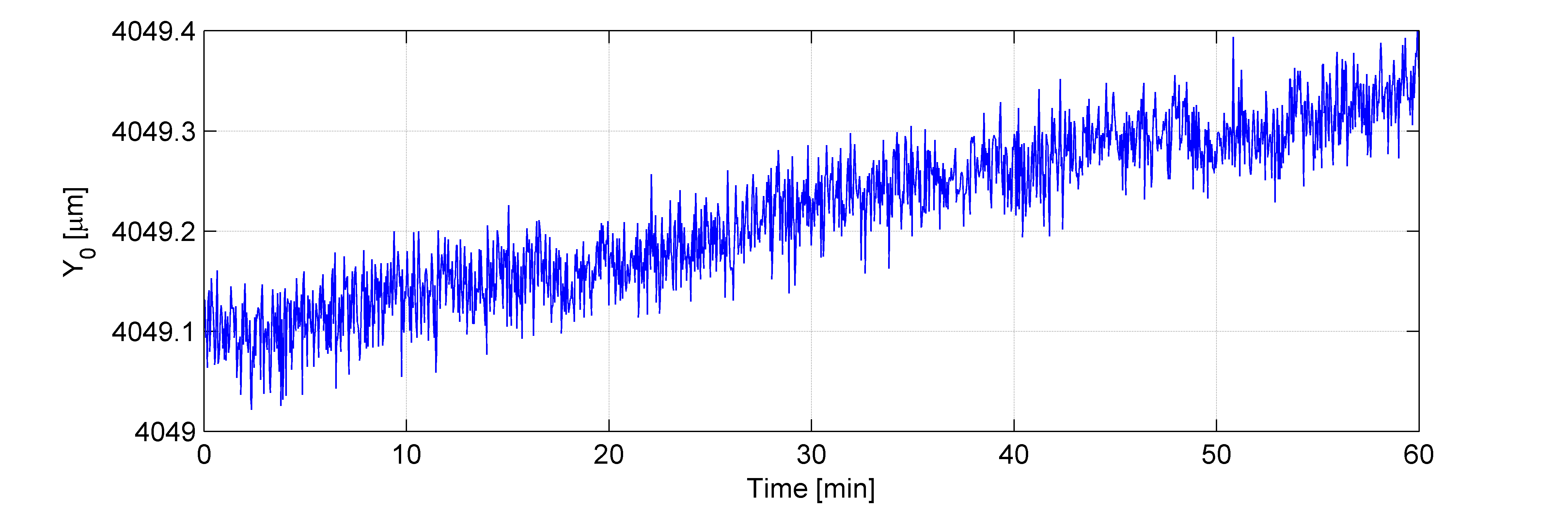}
\caption{The results of a one hour measurement. The center point position $[X_0,Y_0]$ of the laser spot varies only in a range of few tenths of a micrometer.}
\label{f_experiment_long}
\end{figure}

The typical results of a single measurement are shown in Figure \ref{f_experiment_single}. Despite the fine interference fringes and diffraction rings caused by an auxiliary neutral density filter, it is apparent that the observed intensity pattern coincides well with the Gaussian profile calculated by means of nonlinear regression. In contrast to the results of numerical simulations, we found that the errors $\sigma_{X_1}$ and $ \sigma_{Y_1}$ decrease with increasing $L$. Therefore, in the experiment evaluation we generally set $L = 3w_0 = 4.35\,\mathrm{mm}$ providing compromise between accuracy and computational time. The corresponding errors for the example shown in Figure \ref{f_experiment_single} are $\sigma_{X_1} = \sigma_{Y_1} = 0.12\,\mathrm{\mu m}$. 

In Figure~\ref{f_experiment_long} we show the evaluation of a one hour measurement where the spatial and temporal dependence of the laser spot position $[X_0,Y_0]$ is plotted in three graphs. It is apparent that the laser spot position varies only in a range of few tenths of a micrometer. The corresponding errors are $\sigma_{X_2} = 0.074\,\mathrm{\mu m}$ and $\sigma_{Y_2} = 0.079\,\mathrm{\mu m}$. 

The combined measurement errors of the laser spot position for the given example are $\sigma_{X_C} = \sigma_{Y_C} = 0.14\,\mathrm{\mu m}$. The linear measurement range of the image sensor used in the experimentation is only $L - A = 5.25\,\mathrm{mm}$, so the corresponding dynamic range is only 1:38\,000. However, considering the full-frame sensor with the same pixel size but the linear measurement range $L - A = 23.65\,\mathrm{mm}$, we get the dynamic range about 1:170\,000. This is better than the required ratio 1:150\,000. 



In addition, we also investigated the dependence of the errors $\sigma_{X_1}$ and $\sigma_{Y_1}$ on the beam radius $w_0$, which is a function of the measurement distance $d$. In contrast to the numerical simulation, we found that $\sigma_{X_1}$ and $\sigma_{Y_1}$ slightly decrease with decreasing $w_0$. It means that the deviations of the observed and ideal intensity profile are more significant in case of larger beam radius. This finding provides additional flexibility to improve the performance of the measurement method.





\section{Discussion}
\label{res}

The triangulation method with laser beam tracking has proven very suitable for the tilt angle measurement. Considering the full-frame image sensor, we have demonstrated the dynamic range of 1:170\,000. To calculate the angular resolution and the angular measurement range, we take the beam parameters and the measurement distance used in the experimentation. The combined uncertainty of the tilt angle measurement with respect to a single axis of rotation is $\sigma_{\alpha_C} = \sigma_{X_C} / d = 0.13\,\mathrm{\mu rad}$, the corresponding measurement range is $\pm 0.65^{\circ}$. We also determine the upper limit of resolution for the desired angular range $\pm 4^{\circ}$. Neglecting the increase of the dynamic range with the decreasing measurement distance and beam radius, we find the corresponding measurement distance $d = 170\,\mathrm{mm}$ and the angular uncertainty $\sigma_{\alpha_C} = 0.83\,\mathrm{\mu rad}$. This value is still below the desired resolution of 1~microradian. The long-term stability was proven by the statistical evaluation of the overnight measurements divided into hour intervals. We found that the laser spot position varies over this period only in a range of few tenths of a micrometer. In summary, the achieved results fully meet the demands imposed on the tilt angle measurement system.

There are several sources of measurement errors that may significantly affect the results and that should be addressed during the experimentation. The first problem is a possible pointing instability of the laser source which results in a slow angular drift of the output beam. This should be solved by proper selection of a well-stabilized laser device; further significant reduction is possible by means of the temperature stabilization in the laboratory. Another important problem represent mechanical oscillations of the optical setup which can be eliminated by placing a robust base under the optical components. The next source of errors are fluctuations of the refractive index of air caused by the pressure and temperature gradients. Therefore, it is highly advisable to enclose the laser beam path within a protective tube. The special attention should also be paid to the location of the camera which is heated during the measurement and generates a vertical flow of warm air. If this flow crosses the beam path the measurement error may increase by one order or more and significantly reduce the accuracy. It is therefore necessary to spatially separate the camera body from the rest of the measurement system.



\section{Conclusion}
\label{fin}

We have presented a method to measure the angular tilt of the laser guiding mirrors designed for the high-power laser intended for ELI. Our approach was based on the triangulation method with the laser beam tracking. The performance of the method was tested numerically and verified experimentally with satisfactory results that fulfill the demands for the high dynamic range as well as the long-term stability.

\acknowledgments

The authors wish to express thanks for the support of the GACR, project GAP102/10/1813, EU supported projects no CZ.1.05/2.1.00/01.0017, CZ.1.07/2.4.00/31.0016 and CZ.1.07/2.3.00/30.0054 and the research intent RVO:68081731.


\end{document}